\documentclass{article}
\usepackage[english]{babel}
\usepackage[letterpaper,top=2cm,bottom=2cm,left=3cm,right=3cm,marginparwidth=1.75cm]{geometry}
\usepackage{amsmath,amssymb}
\usepackage{graphicx}
\usepackage[colorlinks=true, allcolors=cyan]{hyperref}
\usepackage{lipsum} 
\usepackage{authblk}
\usepackage{url}
\usepackage{hyperref}
\usepackage{color}
\usepackage[normalem]{ulem} 
\usepackage{pbox} 
\usepackage{makecell}
\usepackage{tabularx}
\usepackage{pbox}
\usepackage{tabulary}
\usepackage{array}
\usepackage{rotating}

\providecommand{\keywords}[1]
{
   	
  \textbf{{Keywords:}} #1
}

\title{Optimal control applied to viral competition}

\author[1,2,3]{\underline{Javier L\'opez-Pedrares}}
\author[4,5]{Cristiana J. Silva}
\author[1,3]{M. Elena V\'azquez-Cend\'on}
\author[1,2,\thanks{corresponding author, email: \href{mailto:alberto.perez.munuzuri@usc.es}{alberto.perez.munuzuri@usc.es} }]{Alberto P. Mu\~nuzuri}
\affil[1]{Galician Center for Mathematical Research and Technology (CITMAga), 15782 Santiago de Compostela, Spain}

\affil[2]{Group of Nonlinear Physics, Universidade de Santiago de Compostela, 15782 Santiago de Compostela, Spain}
\affil[3]{Department of Applied Mathematics, Universidade de Santiago de Compostela, 15782 Santiago de Compostela, Spain}
\affil[4]{Department of Mathematics, ISTA, Iscte - Instituto Universitário de Lisboa, 1649-026 Lisbon, Portugal}
\affil[5]{Center for Research and Development in Mathematics and Applications (CIDMA), Department of Mathematics, University of Aveiro, 3810-193 Aveiro, Portugal}

\begin{document}
\date{}
\maketitle
\begin{abstract}

The emergence of mutant lineages within a viral species has become a public health problem, as the existing treatments and drugs are usually more effective on the original lineages than in the mutant ones. The following manuscript presents mathematical models that describe the emergence of these lineages. In order to reduce the damage and possible casualties that can be attributed to these more contagious microorganisms, the theory of optimal control is introduced and a more sophisticated model is proposed to reduce the mutant growth compared to the original one. The analytical study of these models allows us to obtain an overview of the expected behavior over time, which is validated with numerical simulations.

\end{abstract}


\section{Introduction}

Different diseases and pandemics took place over the centuries or even millennia but only in recent years have been studied in depth  \cite{lederberg2000infectious,van2013contagious,piret2021pandemics}. Particularly relevant are all the studies done related to the COVID19 pandemic where the development of strategies that delayed the progress of the sickness where crucial in saving lives \cite{ndairou2020mathematical, khajanchi2021mathematical}. Medical and technological advances allowed scientists to eradicate many diseases and find treatments for others. Advances in mathematics and the creation of mathematical models of infectious diseases also made it possible to consider more complex scenarios \cite{grassly2008mathematical,hethcote2000mathematics} and prepare strategies to deal with them. Many of these models have been validated and their reliability has been demonstrated with empirical data \cite{carballosa2021incorporating}. These models are becoming increasingly complex as pathogens tend to mutate in order to survive over time.  Viruses tend to mutate as well in order to survive within the host organism. Therefore, the infected person at the same instant of time is usually affected by different lineages of the same virus that compete for the same resources.

Two different lineages cannot coexist in an individual in the long run as they are competing for the same resources as follows from the Principle of Competitive Exclusion \cite{hardin1960competitive,levin1970community}. When a mutation occurs and there are two lineages of the same virus competing within the host organism, the best adapted one tends to survive in this process of competition resulting in the extinction of the other and verifying the previous principle. Currently, the study of viral strains has become more complex than previously described. Diseases are more aggressive and we must try to minimize their impact by reducing their spread and number of deaths. Mathematically, this can be represented by including controls in the models and solving optimal control problems. 

Previous literature regarding species competition \cite{wangersky1978lotka,murray2002mathematical,waltman1983competition} and the spread of diseases already exits \cite{siettos2013mathematical,kretzschmar2010mathematical,huppert2013mathematical}.  The phenomenon of competition is usually expressed in terms of a predator-prey mechanism \cite{saez1999dynamics, bunin2017ecological}. A clear example is the study of lynx and hare populations \cite{nedorezov2016dynamics}, but this type of model is also common in fish species \cite{olabanjo2023dynamics}. 
Recent results demonstrated how the connections and mobility of the hosts might influence the evolution of the viruses \cite{lopez2023interactions}. However, these models do not cover viral dispersion thoroughly. Optimal control theory is essential in this type of model, as it allows the inclusion of interventions like medication to mitigate viral effects \cite{sharomi2017optimal,gaff2009optimal}. Historically, in ecosystems, populations were not affected by human impact, except in specific situations such as hunting. Currently, there is an interest in controlling the development of one or several specific species by introducing new variables in our ecosystem \cite{crespo2002optimal}, with the aim of minimizing the environmental cost. There are studies aimed to see how the effect of fishing on a fish predator-prey system can control the system to obtain a totally new dynamic \cite{sager2006numerical}. The same principles will be applied along this manuscript to control a virus sprout.

The article focuses on the effect of introducing a control in classical systems of mutant-variants competition. The control functions describe the effect that an antiviral or vaccine can have on the viral evolution of a disease. We study the new scenario that appears when we introduce commands that reduce viral dispersion in the free competition model. 

The manuscript is organized as follows. In Section~\ref{sec:Model_No_Control}, we present a model of viral competition in a single host infected by a strain and its mutant variant. This section also contains an analytical study of the expected behavior and we carry out simulations showing such behavior. In Section~\ref{sec:Model_Control}, we introduce a control problem associated with the model presented in the previous section. This section, in turn, is structured in two subsections. In the first one, we describe the control problem without state constraints and, in the second one, the problem with the addition of a state constraint. In both cases, numerical simulations are carried out to explain the change introduced by the control in the original model. In addition, a brief analytical study is presented for the case of introducing a constant control in the system. The final section is the discussion and conclusions.

\section{Control-free Competition Model}
\label{sec:Model_No_Control}

A virus can be modeled according to its phenotypic traits. Some fundamental traits can be the reproduction rate, lethality, incubation time or its competitiveness \cite{delong2022towards}.

Successful mutated viruses  change some of the above parameters to improve efficiency in order to survive over time. Let us name the parameters $r$ and $k$ as the reproduction rate and the competition rate, respectively. These are two characteristic parameters of the virus phenotype. The competition parameter, $k$, in our case, will include other characteristics such as lethality. The simplest scheme for species competition is based on the Lotka-Volterra equations \cite{bao2011competitive}.

In the previous study \cite{lopez2023interactions}, we observed the virus phenotype that prevail after competition over time. It was shown that viruses tend to adapt with increasing rates of competition while the reproduction rate, $r$, is not playing any significant role for the system evolution.

A model of viral competition between two species of the same virus, one from the original lineage and another one from the mutant, can be described by the following set of ordinary differential equations \cite{fabre2012modelling},

\begin{equation}
\begin{split}
    \dfrac{dV_A(t)}{dt} = r_A \left( 1 - \dfrac{V_A(t) + V_B(t)}{k_A} \right) V_A(t) \\
    \dfrac{dV_B(t)}{dt} = r_B \left( 1 - \dfrac{V_A(t) + V_B(t)}{k_B} \right) V_B(t),
\end{split}
\label{original}
\end{equation}

\noindent
where $r_A, r_B, k_A, k_B \in \mathbb{R}^+$ and $V_i(t)$ denotes the viral density of species $i$ at time $t$, $r_i$ denotes the reproduction rate of virus $i$, and $k_i$ denotes the competition rate of virus $i$, with $i = A, B$. We consider $k_A < k_B$ to ensure that the second strain is the better-adapted mutant strain, i.e., the more aggressive one. In other words, the one that, over time, is expected to dominate.

The fixed points of the system are calculated as solutions of these equations,

\begin{equation}
\begin{split}
    r_A \left( 1 - \dfrac{V_A(t) + V_B(t)}{k_A} \right) V_A(t) = & \ 0\\
    r_B \left( 1 - \dfrac{V_A(t) + V_B(t)}{k_B} \right) V_B(t) = & \ 0
\end{split}
\end{equation}

To study the stability, we need to compute the Jacobian matrix for the system, as discussed in \cite{strogatz2018nonlinear} and \cite{kuznetsov1998elements}. By substituting the equilibrium points, we obtain the results shown in Table~\ref{table_no_control}. All calculations are provided in detail in the Supplementary material.


\begin{table}[!hbt]
	\caption{Stability analysis of the equilibrium points of the system \eqref{original}.}
	\centering
	
		\begin{tabular}{|c|c|c|}
			\hline
			\textbf{Equilibrium point} & \textbf{Eigenvalues} & \textbf{Stability} \\ 
			\hline
			$(0,0)$ & 
			$\begin{aligned}
				&\lambda_1 = r_A > 0, \\
				&\lambda_2 = r_B > 0
			\end{aligned}$ & 
			Unstable \\ 
			\hline
			$(k_A,0)$ & 
			$\begin{aligned}
				&\lambda_1 = -r_A < 0, \\
				&\lambda_2 = \frac{r_B(-k_A + k_B)}{k_B} > 0
			\end{aligned}$ & 
			Unstable \\ 
			\hline
			$(0,k_B)$ & 
			$\begin{aligned}
				&\lambda_1 = -r_B < 0, \\
				&\lambda_2 = \frac{r_A(k_A - k_B)}{k_A} < 0
			\end{aligned}$ & 
			Stable \\ 
			\hline
	\end{tabular}
	\label{table_no_control}
\end{table}

In order to illustrate the behavior of the above model, numerical simulations have been performed, using a set of parameters $r$ and $k$, which describe the viral phenotype. We consider two viral strains being the second one the more aggressive or lethal, i.e., $k_A < k_B$. 

\begin{figure}[hbt!]
    \centering
    \includegraphics[width=\textwidth]{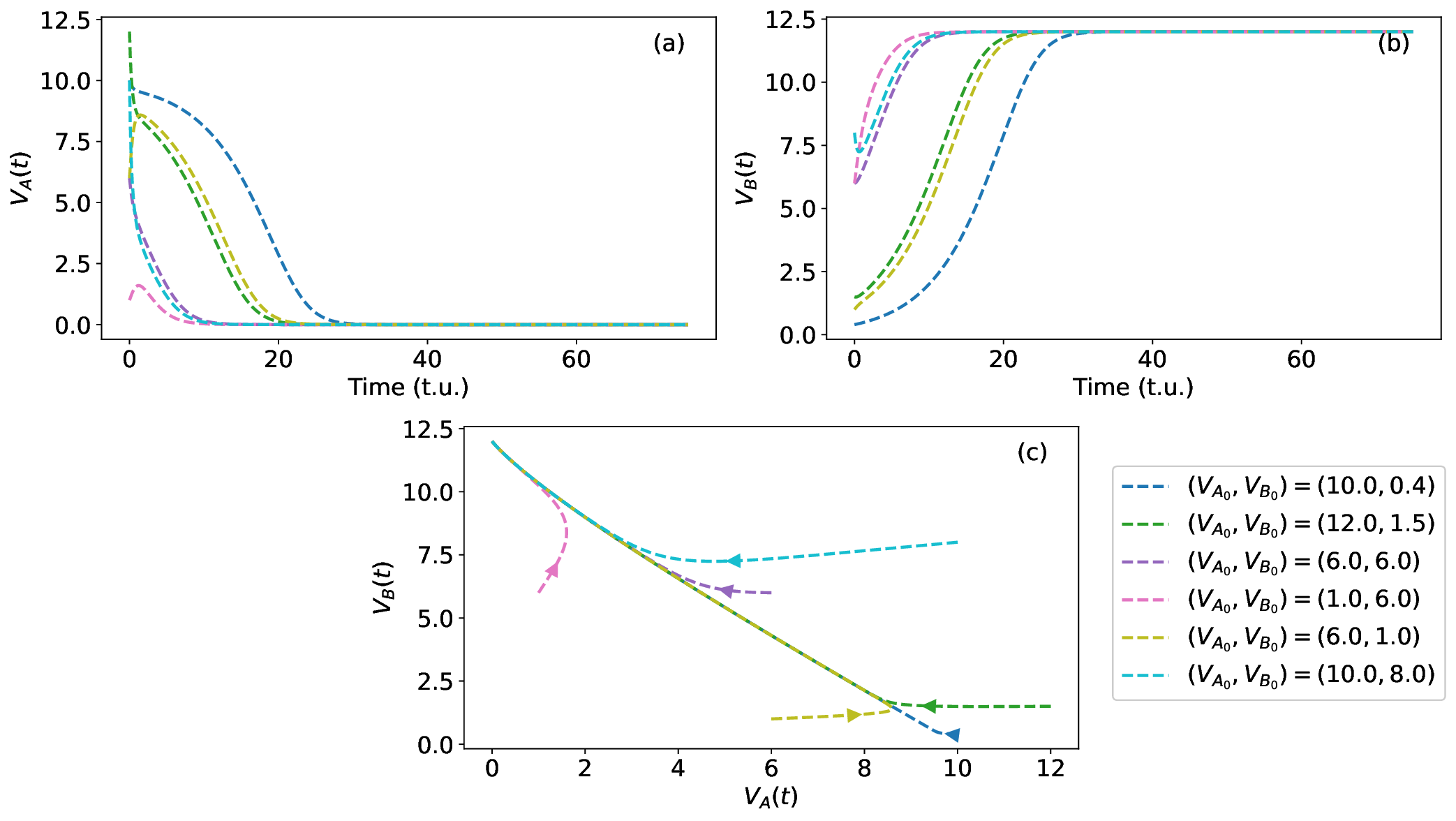}
    \caption{Results for parameter set $r_A = 3$, $r_B = 1$, $k_A = 10$ and $k_B=12$ for different initial conditions. (a) Time evolution of $V_A(t)$, (b) time evolution of $V_B(t)$ and (c) $V_A(t)$ against $V_B(t)$.}
    \label{fig:model_no_control}
\end{figure}

\begin{figure}[ht]
    \centering
    \includegraphics[width=\textwidth]{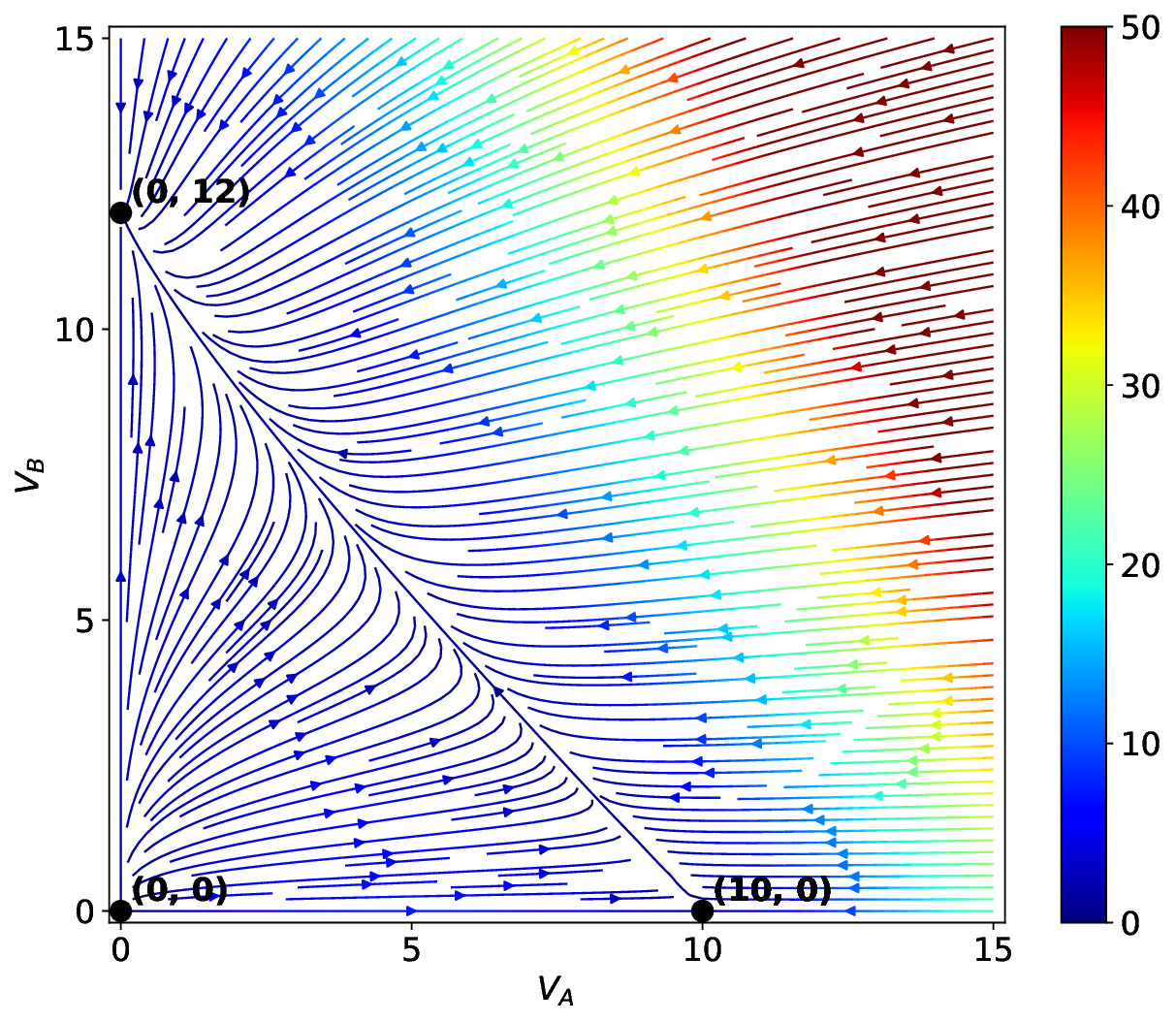}
    \caption{Phase portrait for parameter set $r_A = 3$, $r_B = 1$, $k_A = 10$ and $k_B=12$.}
    \label{fig:phase-portrait}
\end{figure}

The behavior obtained is as expected so that the less aggressive strain dies out and the mutant reaches the carrying capacity of the system as we can see in Figure \ref{fig:model_no_control}. Expressed differently, we reach the stable equilibrium point of the system as seen in Figure \ref{fig:phase-portrait}. Phase portraits and detailed analytical studies of more general systems are detailed in \cite{bazykin1998nonlinear}. Note that we are only interested in solutions containing values for the variables that are positive as they are the only with biological meaning.

Different numerical methods have been used to integrate system \eqref{original}: the trapezium method, explicit Euler and a fourth order Runge-Kutta as shown in Figure \ref{fig:model_no_control}. In all three cases the behavior obtained matches and the differences in the trajectories are negligible. Nevertheless, we must not forget that the numerical methods mentioned are conditionally convergent. Therefore, during the simulations, the time step, $\Delta t$, had to be adjusted to avoid violating the stability conditions of these methods \cite{alexander1990solving,butcher2016numerical,isaacson1994analysis}. To ensure that the system converges, the final simulations were also carried out by integrating the model with an unconditionally convergent method, such as implicit Euler.

\section{Competition model with control}
\label{sec:Model_Control}

Control, in the context of virus infections, has traditionally been made by humans using drugs and vaccines. When a new mutated variant emerges, the conventional treatments are usually less effective and this is the case that we are considering. In the following, we develop strategies that allow the control of the new mutant virus spread by means of the old drugs and treatments. 

In the framework of virology and epidemiology, it is very useful to apply the optimal control theory in order to minimize the impact of a disease \cite{kamyad2014mathematical} as well as to minimize the extend of the intervention. Control can be modeled in this case as the treatment given to a sick person \cite{de2000optimal}. Without applying control we obtain the original model, in which the disease evolves without being treated. When we apply control we are providing treatment to the patient to reduce the viral density present in his organism.

\subsection{Viral optimal control problem}

It should be noted that if we have two lineages of a virus and a single treatment, it is usually more effective for the original variant and less effective for the mutation that has appeared over time. This is why we must consider two constants that regulate the efficacy of the antiviral. Let $c_A$ and $c_B$ be two constants that measure the efficacy against viral strain A and B, respectively. To account for lineage aggressiveness we consider $c_A > c_B$, which means that we have an antiviral that is more effective against the original strain than against the mutant strain.

We consider the system \eqref{original} and introduce a control function $u(\cdot)$, representing the antiviral treatment. Then the dynamics of the control model is given by,

\begin{equation}
    \begin{split}
    \dfrac{dV_A(t)}{dt} = r_A \left( 1 - \dfrac{V_A(t) + V_B(t)}{k_A} \right) V_A(t) - c_A u(t) V_A(t) \\
    \dfrac{dV_B(t)}{dt} = r_B \left( 1 - \dfrac{V_A(t) + V_B(t)}{k_B} \right) V_B(t) - c_B u(t) V_B(t),
    \end{split}
    \label{eq:controlsystem}
\end{equation}

\noindent
where $r_A, r_B, k_A, k_B \in \mathbb{R}^+$, $c_A, c_B \in [0,1]$, additionally $k_A < k_B$ and $c_A > c_B$ and the control function $u(\cdot)$ satisfies the constraint $0 \leq u(\cdot) \leq u_{\max} \leq 1$.

The set of admissible control functions is given by 
\begin{equation*}
	\Omega = \left\{  u(\cdot)  \in L^\infty(0, t_f)  \, \, | \, \,  	0 \leq u(t) \leq u_{\max} \leq 1 \, ,  \quad \forall t \in [0, t_f]  \right\}.
\end{equation*}

When $u(t) = 0$ no treatment is supplied to the system, i.e., we recover the original system proposed in \eqref{original}. On the other hand, when $u(t) \in (0,1]$ we are introducing antiviral to the system, which reduces the spread of the disease. Moreover, we consider constant initial conditions   $\left(V_A(0) , V_B(0)\right) =\left( V_{A_{0}},V_{B_{0}} \right) \in \mathbb{R}^+ \times \mathbb{R}^+$.

The main goal is to minimize the mutant strain $V_B(t)$, in the time window $[0,t_f]$, i.e., to approach the fixed point of the original system \eqref{original}, with the less cost possible, therefore and following  \cite{ibanez2017optimal}, we consider the following cost functional, $J$,
\begin{equation}
J\left(V_A(t),V_B(t),u(t)\right) =  \displaystyle\int_0^{t_f}  \left[\left( V_A(t) - k_A \right)^2 + \left( V_B(t) - 0 \right)^2 + u^2(t) \right]\, dt .
\label{eq:functional}
\end{equation}

This functional is defined in order to approximate the viral density $V_A$ to $k_A$ and $V_B$ to extinction. In other words, we try to minimize the cost of approaching the unstable point of the original system \eqref{original}.

\subsubsection{Analytical study of the problem with constant control}

The equilibrium points in the  $V_A$-$V_B$ plane result modified once the control is introduced. In this subsection we study the behavior of the system if we apply a constant dose of antiviral over time, i.e., when a constant value is considered for the control, $u$. They are obtained from,



\begin{equation}
\begin{split}
    r_A \left( 1 - \dfrac{V_A(t) + V_B(t)}{k_A} \right) V_A(t) - c_A u V_A(t) = 0\\
    r_B \left( 1 - \dfrac{V_A(t) + V_B(t)}{k_B} \right) V_B(t) - c_B u V_B(t) = 0,
\end{split}
\label{constant_control}
\end{equation}

\noindent 
which is equivalent to solving the above linear system,

\begin{equation}
\begin{split}
    r_A \left( 1 - \dfrac{V_A(t) + V_B(t)}{k_A} \right)  - c_A u  = 0\\
    r_B \left( 1 - \dfrac{V_A(t) + V_B(t)}{k_B} \right)  - c_B u = 0.
\end{split}
\end{equation}

The fixed points are the solutions for the previous coupled equations. To study their stability we need to compute the Jacobian matrix for the system. 
As in the previous section, we substitute the equilibrium points and obtain the results shown in Table~\eqref{cond_control}. The necessary calculations are provided in detail in the Supplementary material.

\begin{table}[!hbt]
	\caption{Stability conditions for the equilibrium points of the system \eqref{constant_control}.}
	\centering
	\resizebox{\textwidth}{!}{
		\begin{tabular}{|c|c|c|}
			\hline
			\textbf{Equilibrium point} & \textbf{Eigenvalues} & \textbf{Stability conditions} \\ 
			\hline
			$(0,0)$ & 
			$\begin{aligned}
				&\lambda_1 = r_A - c_A u \\
				&\lambda_2 = r_B - c_B u
			\end{aligned}$ & 
			$\max\left\{\frac{r_A}{c_A},\frac{r_B}{c_B}\right\} < u$ \\ 
			\hline
			$\left(-k_A \left( \dfrac{c_A u - r_A}{r_A} \right), 0 \right)$ &  
			$\begin{aligned}
				&\lambda_1 = c_A u - r_A \\
				&\lambda_2 = r_B \left(1 + \dfrac{k_A}{r_A k_B}(c_A u - r_A)\right) - c_B u
			\end{aligned}$ & 
			$\begin{aligned}
				&r_B \left(1 - \dfrac{k_A}{k_B}\right) < \left(c_B - \dfrac{r_B k_A}{r_A k_B} c_A\right) u \\
				&u < \dfrac{r_A}{c_A}
			\end{aligned}$ \\ 
			\hline
			$\left(0, -k_B \left( \dfrac{c_B u - r_B}{r_B} \right)\right)$ & 
			$\begin{aligned}
				&\lambda_1 = r_A \left( 1 + \dfrac{k_B}{r_B k_A}(c_B u - r_B)\right) - c_A u \\
				&\lambda_2 = c_B u - r_B
			\end{aligned}$ & 
			$\begin{aligned}
				&r_A \left(1 - \dfrac{k_B}{k_A}\right) < \left(c_A - \dfrac{r_A k_B}{r_B k_A} c_B\right) u \\
				&u < \dfrac{r_B}{c_B}
			\end{aligned}$ \\
			\hline
			$V_B^* = - V_A^* + k_A - \dfrac{k_A c_A}{r_A} u$ & 
			$\begin{aligned}
				&\lambda_1 = 0 \\
				&\lambda_2 = \dfrac{r_A}{k_A} V_A^* + \dfrac{r_B}{k_B} V_B^* > 0
			\end{aligned}$ & 
			Unstable \\ 
			\hline
	\end{tabular}}
	\label{cond_control}
\end{table}

The previous analytical study that has been carried out allows characterizing the equilibrium points of the system with constant control. In addition, it establishes a relationship between the system parameters that allows us to know when the mutant strain, $V_B$, becomes extinct.


After calculating the equilibrium points and studying their stability for the problem with constant control, numerical simulations have been carried out in order to observe the behavior for a given set of parameters $r$, $k$, $c$ and $u$. Again as before, different numerical methods have been used to integrate the system.

\begin{figure}[ht]
    \centering
    \includegraphics[width=\textwidth]{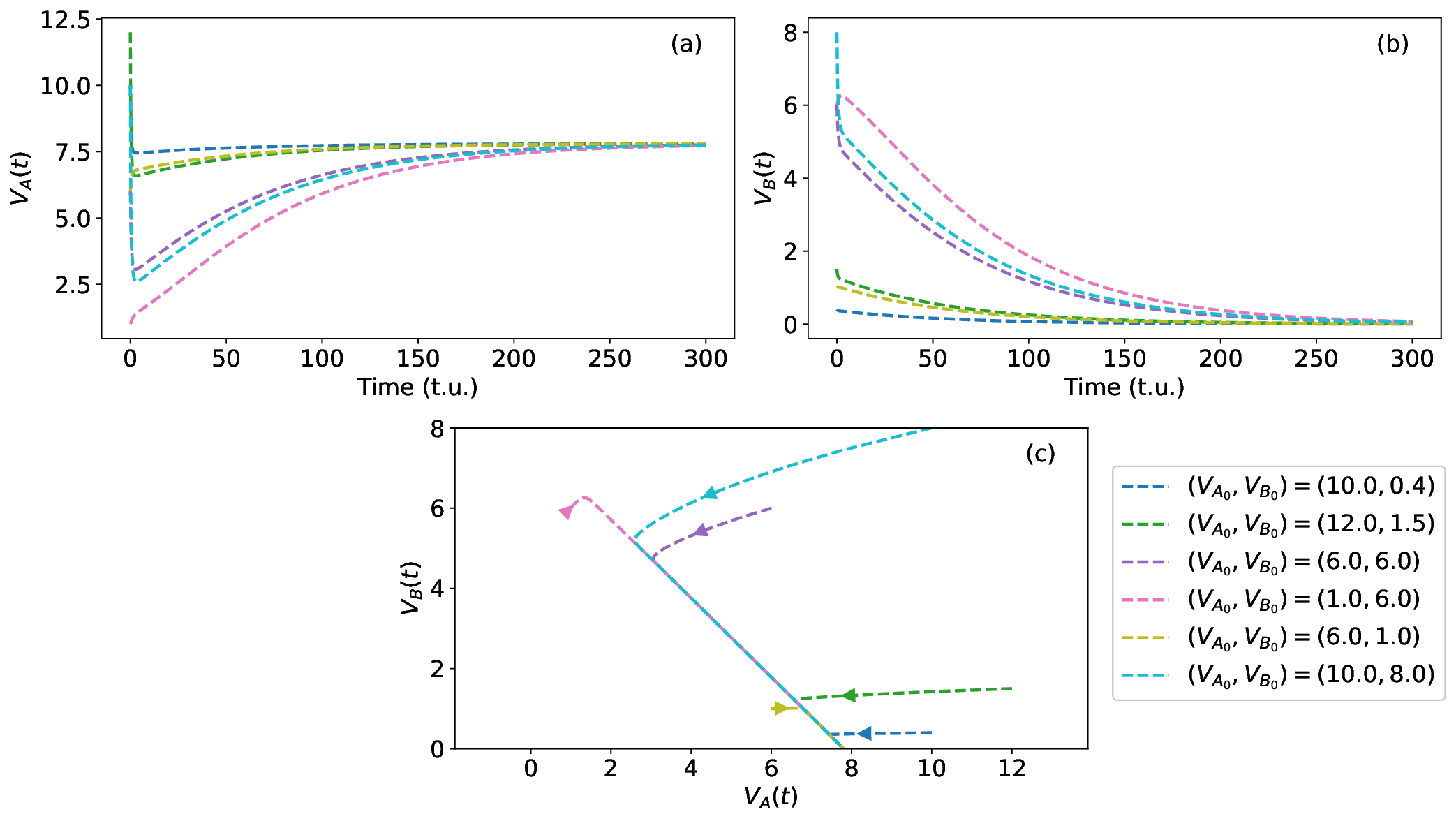}
    \caption{Results for parameter set $r_A = 3$, $r_B = 1$, $k_A = 10$, $k_B=12$, $c_A = 0.9$, $c_B = 0.5$ and constant control $u = 0.733097$ for different initial conditions. (a) Time evolution of $V_A(t)$, (b) time evolution of $V_B(t)$ and (c) $V_A(t)$ against $V_B(t)$.}
    \label{fig:model_constant_control}
\end{figure}

\begin{figure}[ht]
    \centering
    \includegraphics[width=\textwidth]{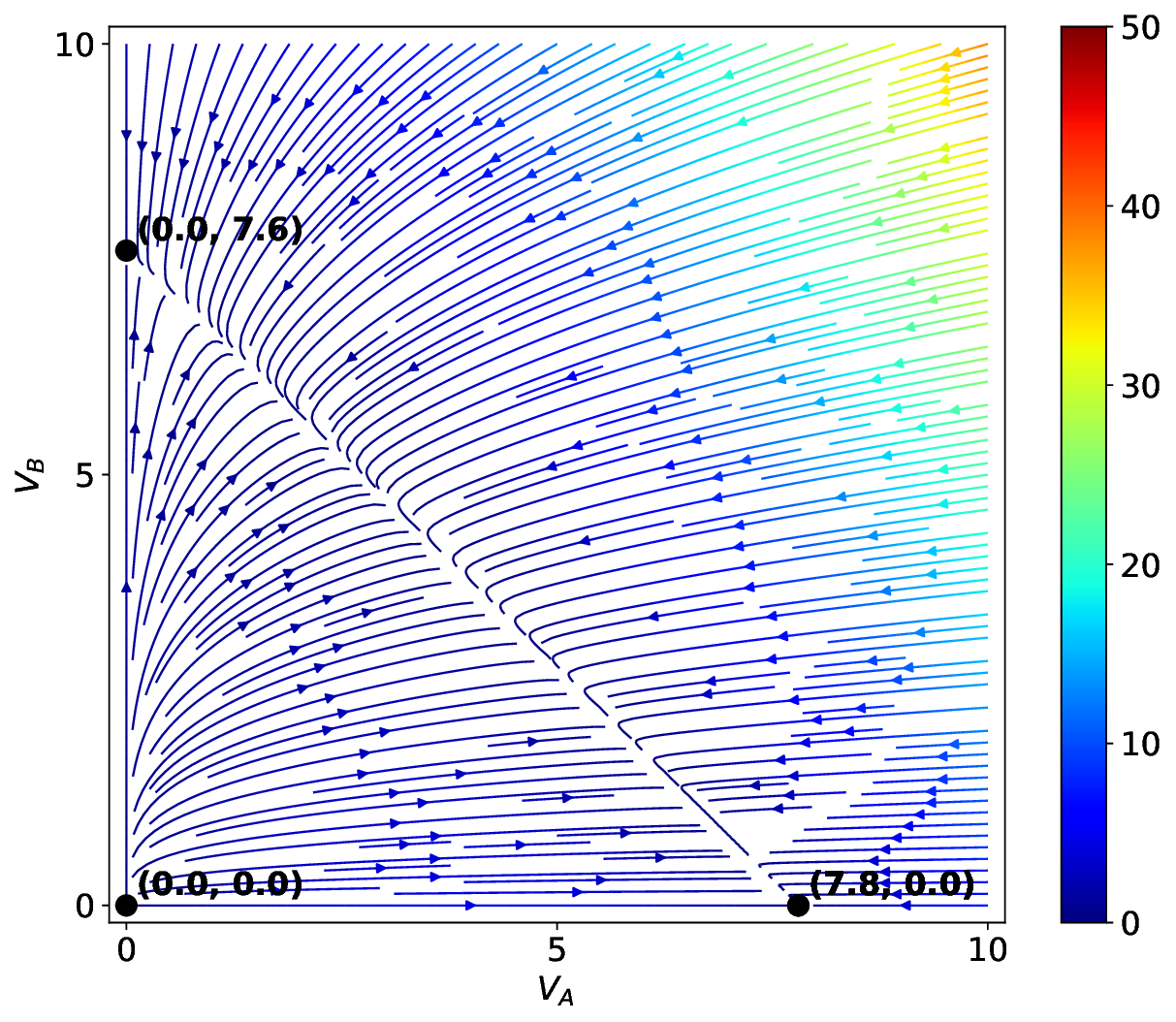}
    \caption{Phase portrait for parameter set $r_A = 3$, $r_B = 1$, $k_A = 10$, $k_B=12$, $c_A = 0.9$, $c_B = 0.5$ and constant control $u = 0.733097$ for different initial conditions.}
    \label{fig:phase-portrait-constant-control}
\end{figure}

As we can see in Figure \ref{fig:model_constant_control} for the selection of suitable parameters, with constant control, we are able to invert the dynamics of the original system. It is important to note that in this case the system does not reach equilibrium in such a short time as in the case without medical treatment (i.e. control). Now, to reach the equilibrium state we must integrate the system four times as long. In this case, we observe that the predominant viral strain in the system at the end of the simulation is the least aggressive, $V_A$. In other words, by applying a constant drug dose, we are able to reach in the new model the equilibrium point equivalent to the unstable one in the original model as we can see in Figure \ref{fig:phase-portrait-constant-control}. It is easy to verify that the chosen parameter set verifies the stability conditions.

\subsubsection{Numerical simulations for the optimal control problem using a direct method}

Numerical simulations have been carried out using \textit{OptimalControl toolbox} \cite{Caillau_OptimalControl_jl_a_Julia} implemented in \textit{Julia} programming language.

 The implementation of the problem in \textit{Julia} can be complex, so we write the optimal control problem considering a new state variable, $V_C$, related to the cost function to reduce the difficulty and our optimal control problem, then, becomes,

\begin{equation}
    \tag{OCP}
    \begin{cases}
        V_C(t_f) + \displaystyle\int_0^{t_f} u^2(t) \ dt \rightarrow \text{min} \\[0.2cm]
        \dot{V}_A(t) = r_A \left( 1 - \dfrac{V_A(t) + V_B(t)}{k_A} \right) V_A(t) - c_A u(t) V_A(t) \\
        \dot{V}_B(t) = r_B \left( 1 - \dfrac{V_A(t) + V_B(t)}{k_B} \right) V_B(t) - c_B u(t) V_B(t) \\ \dot{V}_C(t) =\left( V_A(t) - k_A \right)^2 + \left( V_B(t) - 0 \right)^2
        \\[0.2cm]
        u \in [0, u_{\text{max}}], \ t \in [t_0 , t_f] \ \text{a.e.},\\
        V_A(0) = V_{A_{0}}, \ V_B(0) = V_{B_{0}}, \ V_C(0) = 0 \, .
    \end{cases}
    \label{OCP}
\end{equation}

\begin{figure}[ht]
    \centering
    \includegraphics[width=\textwidth]{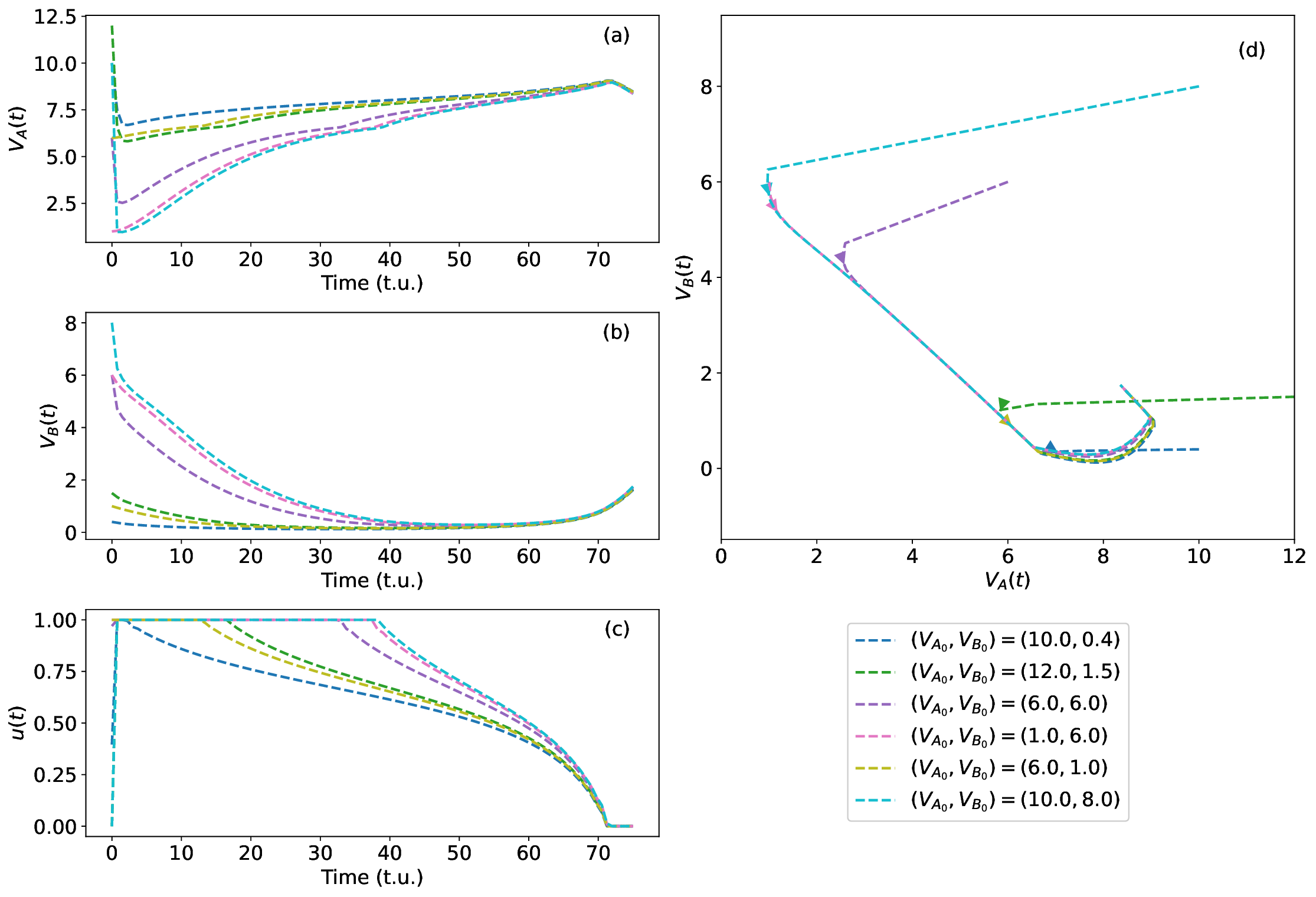}
    \caption{Results for parameter set $r_A = 3$, $r_B = 1$, $k_A = 10$, $k_B=12$, $c_A=0.9$ and $c_B =0.5$ for different initial conditions. (a) Time evolution of $V_A(t)$, (b) time evolution of $V_B(t)$, (c) time evolution of $u(t)$ and (d) $V_A(t)$ against $V_B(t)$.}
    \label{fig:model_control}
\end{figure}

As we can observe in Figure \ref{fig:model_control}, u(t) allows to control the mutant lineage so that it never outperforms the original variant in the simulation time.

Note that, in this case, the system tries to approach the equilibrium point, $(k_A,0)$, which was initially unstable for the chosen set of parameters in the original system without control \eqref{original}. However, this point is never fully reached and once the control is removed $V_B(t)$ grows exponentially (see Figure \ref{fig:model_control} for times larger than 60 t.u.).


\subsection{Viral optimal control problem with state and control constraints}

As described in the previous subsection, the problem eventually tends to an exponential growth of the mutant strain.  The initial goal was not only to make sure that $V_A(t) \geq V_B(t)$ for all $t \in [0, t_f]$, but also that the mutant strain, $V_B(t)$, did not grow and was kept as small as possible.
For that, a new constraint is included in 
one of the state variables of the optimal control problem. If we want the most aggressive variant, $V_B(t)$, to remain below a threshold close to zero, then, we must consider the following constraint,

\begin{equation*}
   0 \leq V_B(t) \leq \xi\ \quad  \forall t \in [0, t_f],
\end{equation*}

\noindent
with $\xi >0$ and close to zero.

We consider, thus, the following optimal control problem with state and control constraints,

\begin{equation}
\tag{OCP-WSC}
\begin{cases}
V_C(t_f) + \displaystyle\int_0^{t_f} u^2(t) \ dt \rightarrow \text{min} \\[0.2cm]
\dot{V}_A(t) = r_A \left( 1 - \dfrac{V_A(t) + V_B(t)}{k_A} \right) V_A(t) - c_A u(t) V_A(t) \\
    \dot{V}_B(t) = r_B \left( 1 - \dfrac{V_A(t) + V_B(t)}{k_B} \right) V_B(t) - c_B u(t) V_B(t) \\ \dot{V}_C(t) =\left( V_A(t) - k_A \right)^2 + \left( V_B(t) - 0 \right)^2
\\[0.2cm]
0 \leq V_B(t) \leq \xi, \
u \in [0, u_{\text{max}}], \ t \in [t_0 , t_f] \ \text{a.e.},\\
V_A(0) = V_{A_{0}}, \ V_B(0) = V_{B_{0}}, \ V_C(0) = 0 \, ,
\end{cases}
\label{OCP-WSC}
\end{equation}

\noindent
where the variables and parameters are the same as in the previous models.

\subsubsection{Numerical simulations for the \texorpdfstring{\eqref{OCP-WSC}}{OCP-WSC} using a direct method}

\begin{figure}[ht]
    \centering
    \includegraphics[width=\textwidth]{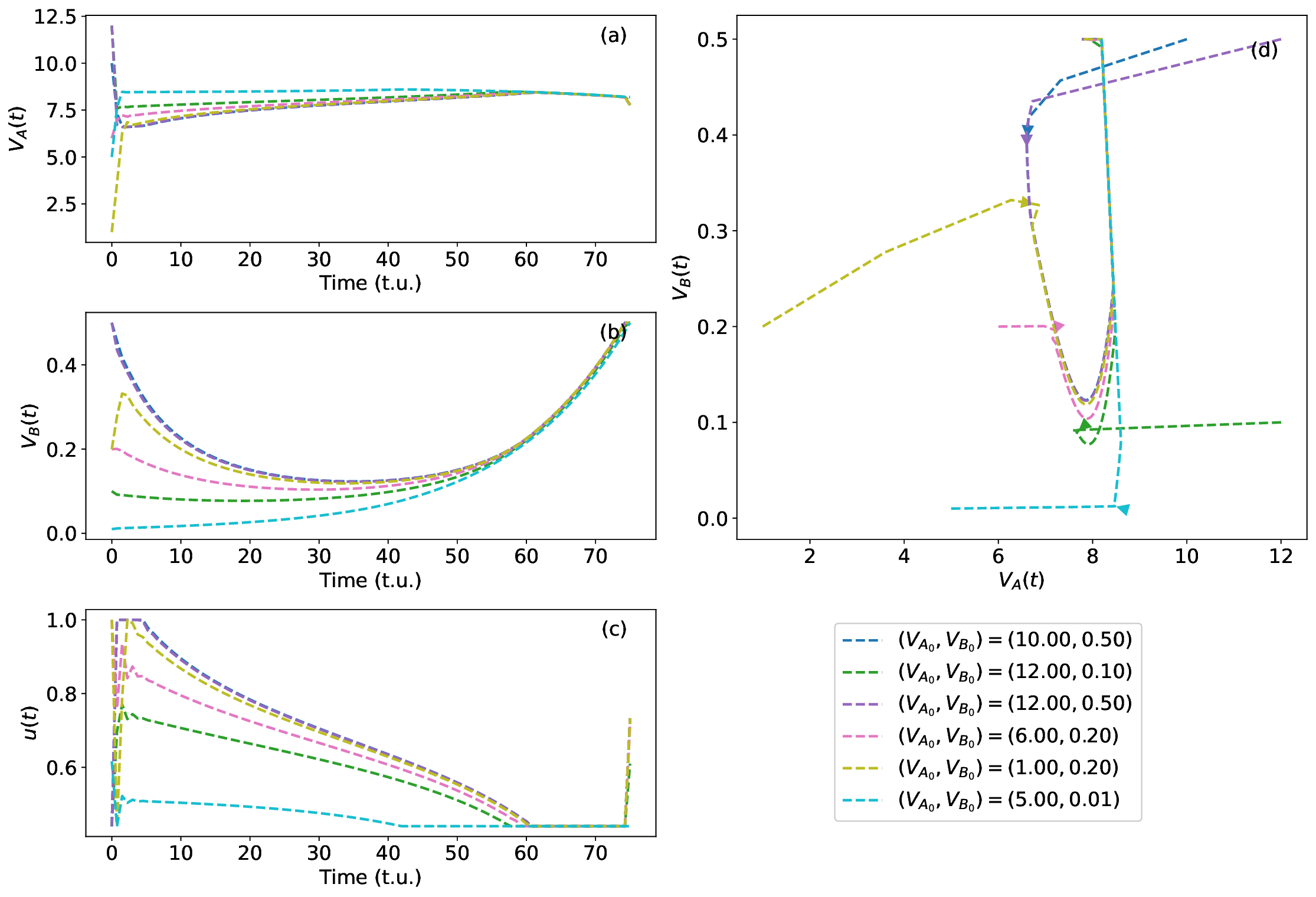}
    \caption{Results for parameter set $r_A = 3$, $r_B = 1$, $k_A = 10$ and $k_B=12$ for different initial conditions. (a) Time evolution of $V_A(t)$, (b) time evolution of $V_B(t)$, (c) time evolution of $u(t)$ and (d) $V_A(t)$ against $V_B(t)$.}
    \label{fig:model_control_constraint}
\end{figure}

As before, using the \textit{OptimalControl toolbox} implemented in \textit{Julia}, we can solve the control problem with constraints on the state variable associated with the mutant lineage. Similar to the behavior observed in Figure \ref{fig:model_control} for the \eqref{OCP} problem in which we had no constraints on the state, we observe in Figure \ref{fig:model_control_constraint} the growth of the original versus the mutant variant associated with the \eqref{OCP-WSC} problem. If we look at the behavior of the mutant strain, $V_B$, in Figure \ref{fig:model_control_constraint}, over time we have managed to keep it below a threshold set close to zero, $\xi$. This implies that, while we can manage the emerging variant through treatment, the optimal control approach does not achieve eradication. Rather, it identifies the most cost-effective treatment strategy to control the strains and their anticipated temporal distribution for medication administration.

\section{Conclusions}

Throughout the manuscript, a simple model of viral competition between two variants of the same virus, the original strain and its mutation, has been developed and implemented.  In this context, we have simultaneously introduced a treatment for the disease utilizing optimal control theory. We focused on the dynamics and the previous analytical study of the system without control and with control, in order to observe how introducing control allowed us to completely modify the original dynamics. The described models verify one of the basic principles of biology, the Principle of Competitive Exclusion, which states that a single virus survives when there are several competing for the same resources. In our case, we choose that the mutant lineage is the one that survives in the absence of the control and with treatment we are able to reverse the behavior. The effect of control, that within this context means the administration of antivirals,
in the models influences in two ways. First, introducing an appropriate control allows the original system to exchange dynamics. Second, choosing the optimal control allows with the lowest-cost treatment  to minimize the spread of a more aggressive lineage. It is also important to note that the control introduced allows in all cases considered to delay the outburst of the mutant variant and, thus, delay its harmful effects on the hosts.

In conclusion, introducing a control variable, a treatment, changes the mechanisms of viral competition and makes it possible to obtain lineages that are less harmful to the potential hosts.

\section*{Declaration of competing interest}
The authors declare that they have no known competing financial interests or personal relationships that could have appeared to influence the work reported in this paper.

\section*{Acknowledgements}
APM and JLP gratefully acknowledge financial support by the Spanish Ministerio de Economía y Competitividad and European Regional Development Fund under contract PID 2020-113881RB-I00 AEI/FEDER, UE, and by Xunta de Galicia under Research Grant No. 2021-PG036. All these programs are co-funded by FEDER (UE). 
EVC received financial support from the Xunta de Galicia (2021 GRC GI-1563 - ED431C 2021/15)
and this work has been partially supported by FEDER, Ministerio de Ciencia e Innovación-AEI research project PID2021-122625OB-I00.
CJS was partially supported by the Portuguese Foundation for Science and Technology (FCT), through the Center for Research and Development in Mathematics and Applications (CIDMA), projects UIDB/04106/2020 and UIDP/04106/2020, and within the project “Mathematical Modelling of Multiscale Control Systems: Applications to Human Diseases” (CoSysM3), Reference 2022.03091.PTDC, financially supported by national funds (OE) through FCT/MCTES.
The simulations were run in the Supercomputer Center of Galicia (CESGA) and we acknowledge their support.

\bibliographystyle{unsrt}
\bibliography{sample}

\begin{thebibliography}{10}

\bibitem{lederberg2000infectious}
Joshua Lederberg.
\newblock Infectious history.
\newblock {\em Science}, 288(5464):287--293, 2000.

\bibitem{van2013contagious}
Willem~G Van~Panhuis, John Grefenstette, Su~Yon Jung, Nian~Shong Chok, Anne
  Cross, Heather Eng, Bruce~Y Lee, Vladimir Zadorozhny, Shawn Brown, Derek
  Cummings, et~al.
\newblock Contagious diseases in the united states from 1888 to the present,
  2013.

\bibitem{piret2021pandemics}
Jocelyne Piret and Guy Boivin.
\newblock Pandemics throughout history.
\newblock {\em Frontiers in microbiology}, 11:631736, 2021.

\bibitem{ndairou2020mathematical}
Fa{\"\i}{\c{c}}al Nda{\"\i}rou, Iv{\'a}n Area, Juan~J Nieto, and Delfim~FM
  Torres.
\newblock Mathematical modeling of covid-19 transmission dynamics with a case
  study of wuhan.
\newblock {\em Chaos, Solitons \& Fractals}, 135:109846, 2020.

\bibitem{khajanchi2021mathematical}
Subhas Khajanchi, Kankan Sarkar, Jayanta Mondal, Kottakkaran~Sooppy Nisar, and
  Sayed~F Abdelwahab.
\newblock Mathematical modeling of the covid-19 pandemic with intervention
  strategies.
\newblock {\em Results in Physics}, 25:104285, 2021.

\bibitem{grassly2008mathematical}
Nicholas~C Grassly and Christophe Fraser.
\newblock Mathematical models of infectious disease transmission.
\newblock {\em Nature Reviews Microbiology}, 6(6):477--487, 2008.

\bibitem{hethcote2000mathematics}
Herbert~W Hethcote.
\newblock The mathematics of infectious diseases.
\newblock {\em SIAM review}, 42(4):599--653, 2000.

\bibitem{carballosa2021incorporating}
Alejandro Carballosa, Mariamo Mussa-Juane, and Alberto~P Munuzuri.
\newblock Incorporating social opinion in the evolution of an epidemic spread.
\newblock {\em Scientific reports}, 11(1):1772, 2021.

\bibitem{hardin1960competitive}
Garrett Hardin.
\newblock The competitive exclusion principle: an idea that took a century to
  be born has implications in ecology, economics, and genetics.
\newblock {\em science}, 131(3409):1292--1297, 1960.

\bibitem{levin1970community}
Simon~A Levin.
\newblock Community equilibria and stability, and an extension of the
  competitive exclusion principle.
\newblock {\em The American Naturalist}, 104(939):413--423, 1970.

\bibitem{wangersky1978lotka}
Peter~J Wangersky.
\newblock Lotka-volterra population models.
\newblock {\em Annual Review of Ecology and Systematics}, 9:189--218, 1978.

\bibitem{murray2002mathematical}
James~D Murray.
\newblock Mathematical biology: I. an introduction. interdisciplinary applied
  mathematics.
\newblock {\em Mathematical Biology, Springer}, 17, 2002.

\bibitem{waltman1983competition}
Paul Waltman.
\newblock {\em Competition models in population biology}.
\newblock SIAM, 1983.

\bibitem{siettos2013mathematical}
Constantinos~I Siettos and Lucia Russo.
\newblock Mathematical modeling of infectious disease dynamics.
\newblock {\em Virulence}, 4(4):295--306, 2013.

\bibitem{kretzschmar2010mathematical}
Mirjam Kretzschmar and Jacco Wallinga.
\newblock Mathematical models in infectious disease epidemiology.
\newblock {\em Modern infectious disease epidemiology: Concepts, methods,
  mathematical models, and public health}, pages 209--221, 2010.

\bibitem{huppert2013mathematical}
Amit Huppert and Guy Katriel.
\newblock Mathematical modelling and prediction in infectious disease
  epidemiology.
\newblock {\em Clinical microbiology and infection}, 19(11):999--1005, 2013.

\bibitem{saez1999dynamics}
Eduardo S{\'a}ez and Eduardo Gonz{\'a}lez-Olivares.
\newblock Dynamics of a predator-prey model.
\newblock {\em SIAM Journal on Applied Mathematics}, 59(5):1867--1878, 1999.

\bibitem{bunin2017ecological}
Guy Bunin.
\newblock Ecological communities with lotka-volterra dynamics.
\newblock {\em Physical Review E}, 95(4):042414, 2017.

\bibitem{nedorezov2016dynamics}
LV~Nedorezov.
\newblock The dynamics of the lynx--hare system: an application of the
  lotka--volterra model.
\newblock {\em Biophysics}, 61(1):149--154, 2016.

\bibitem{olabanjo2023dynamics}
Olusola Olabanjo and Ashiribo Wusu.
\newblock Dynamics of sustainable fisheries: a mathematical approach using
  lotka-volterra equations.
\newblock {\em Annals of Mathematics and Computer Science}, 19:59--67, 2023.

\bibitem{lopez2023interactions}
Javier L{\'o}pez-Pedrares, M~Elena V{\'a}zquez-Cend{\'o}n, and Alberto~P
  Mu{\~n}uzuri.
\newblock Interactions between hosts affect virus competition mechanism within
  an infectious strain.
\newblock {\em Chaos, Solitons \& Fractals}, 170:113344, 2023.

\bibitem{sharomi2017optimal}
Oluwaseun Sharomi and Tufail Malik.
\newblock Optimal control in epidemiology.
\newblock {\em Annals of Operations Research}, 251:55--71, 2017.

\bibitem{gaff2009optimal}
Holly Gaff and Elsa Schaefer.
\newblock Optimal control applied to vaccination and treatmentstrategies for
  various epidemiological models.
\newblock {\em Mathematical biosciences \& engineering}, 6(3):469--492, 2009.

\bibitem{crespo2002optimal}
LG~Crespo and JQ~Sun.
\newblock Optimal control of populations of competing species.
\newblock {\em Nonlinear Dynamics}, 27:197--210, 2002.

\bibitem{sager2006numerical}
Sebastian Sager, Hans~Georg Bock, Moritz Diehl, Gerhard Reinelt, and Johannes~P
  Schloder.
\newblock Numerical methods for optimal control with binary control functions
  applied to a lotka-volterra type fishing problem.
\newblock In {\em Recent Advances in Optimization}, pages 269--289. Springer,
  2006.

\bibitem{delong2022towards}
John~P DeLong, Maitham~A Al-Sammak, Zeina~T Al-Ameeli, David~D Dunigan, Kyle~F
  Edwards, Jeffry~J Fuhrmann, Jason~P Gleghorn, Hanqun Li, Kona Haramoto,
  Amelia~O Harrison, et~al.
\newblock Towards an integrative view of virus phenotypes.
\newblock {\em Nature Reviews Microbiology}, 20(2):83--94, 2022.

\bibitem{bao2011competitive}
Jianhai Bao, Xuerong Mao, Geroge Yin, and Chenggui Yuan.
\newblock Competitive lotka--volterra population dynamics with jumps.
\newblock {\em Nonlinear Analysis: Theory, Methods \& Applications},
  74(17):6601--6616, 2011.

\bibitem{fabre2012modelling}
Fr{\'e}d{\'e}ric Fabre, Josselin Montarry, J{\'e}r{\^o}me Coville, Rachid
  Senoussi, Vincent Simon, and Benoit Moury.
\newblock Modelling the evolutionary dynamics of viruses within their hosts: a
  case study using high-throughput sequencing.
\newblock {\em PLoS Pathogens}, 8(4):e1002654, 2012.

\bibitem{strogatz2018nonlinear}
Steven~H Strogatz.
\newblock {\em Nonlinear dynamics and chaos: with applications to physics,
  biology, chemistry, and engineering}.
\newblock CRC press, 2018.

\bibitem{kuznetsov1998elements}
Yuri~A Kuznetsov, Iu~A Kuznetsov, and Y~Kuznetsov.
\newblock {\em Elements of applied bifurcation theory}, volume 112.
\newblock Springer, 1998.

\bibitem{bazykin1998nonlinear}
Alexander~D Bazykin.
\newblock {\em Nonlinear dynamics of interacting populations}.
\newblock World Scientific, 1998.

\bibitem{alexander1990solving}
Roger Alexander.
\newblock Solving ordinary differential equations i: Nonstiff problems (e.
  hairer, sp norsett, and g. wanner).
\newblock {\em Siam Review}, 32(3):485, 1990.

\bibitem{butcher2016numerical}
John~Charles Butcher.
\newblock {\em Numerical methods for ordinary differential equations}.
\newblock John Wiley \& Sons, 2016.

\bibitem{isaacson1994analysis}
Eugene Isaacson and Herbert~Bishop Keller.
\newblock {\em Analysis of numerical methods}.
\newblock Courier Corporation, 1994.

\bibitem{kamyad2014mathematical}
Ali~Vahidian Kamyad, Reza Akbari, Ali~Akbar Heydari, and Aghileh Heydari.
\newblock Mathematical modeling of transmission dynamics and optimal control of
  vaccination and treatment for hepatitis b virus.
\newblock {\em Computational and mathematical methods in medicine},
  2014(1):475451, 2014.

\bibitem{de2000optimal}
JAM~Felippe De~Souza, Marco Antonio~Leonel Caetano, and Takashi Yoneyama.
\newblock Optimal control theory applied to the anti-viral treatment of aids.
\newblock In {\em Proceedings of the 39th IEEE conference on decision and
  control (Cat. No. 00CH37187)}, volume~5, pages 4839--4844. IEEE, 2000.

\bibitem{ibanez2017optimal}
Aitziber Iba{\~n}ez.
\newblock Optimal control of the lotka--volterra system: turnpike property and
  numerical simulations.
\newblock {\em Journal of biological dynamics}, 11(1):25--41, 2017.

\bibitem{Caillau_OptimalControl_jl_a_Julia}
Jean-Baptiste Caillau, Olivier Cots, Joseph Gergaud, Pierre Martinon, and
  Sophia Sed.
\newblock {OptimalControl.jl: a Julia package to model and solve optimal
  control problems with ODE's}.

\end{thebibliography}

\end{document}